\documentstyle[aps]{revtex}
\begin{document}
\twocolumn[\hsize\textwidth\columnwidth\hsize\csname
@twocolumnfalse\endcsname

\title{{\Large \bf A note on the foundation of relativistic
mechanics}\\[2mm] II: Covariant hamiltonian general 
relativity}

\author{Carlo Rovelli \\
{\it Centre de Physique Th\'eorique, Luminy, F-13288 Marseille, EU}\\
{\it Physics Department, Pittsburgh
University, PA-15260, USA}}
\date{\today}
\maketitle    
\begin{abstract} 
I illustrate a simple hamiltonian formulation of general relativity,
derived from the work of Esposito, Gionti and Stornaiolo, which is
manifestly 4d generally covariant and is defined over a finite
dimensional space.  The spacetime coordinates drop out of the
formalism, reflecting the fact that they are not related to
observability.  The formulation can be interpreted in terms of
Toller's reference system transformations, and provides a physical
interpretation for the spinnetwork to spinnetwork transition
amplitudes computable in principle in loop quantum gravity and in the
spin foam models.

\end{abstract}
\vskip1cm] 

\section{The problem}    

In the companion paper \cite{c1} I have discussed the possibility of a
relativistic foundation of mechanics and I have argued that the usual
notions of state and observable have to be modified in order to work
well in a relativistic context.  Here I apply this point of view to
field theory.  In the context of field theory, the relativistic notion
of observable suggests to formulate the hamiltonian theory over a
finite dimensional space, for two reasons.  First the space of the
relativistic (``partial" \cite{r}) observables of a field theory is
finite dimensional.  Second, the infinite dimensional space of the
initial values of the fields, which is the conventional arena for
hamiltonian field theory, is based on the notion of instantaneity
surface, which has little general significance in a relativistic
context.  The possibility of defining hamiltonian field theory over a
finite dimensional space has been explored by several authors (See
\cite{JK,FTHM,HK}, and ample references therein), developing the
classic works of Weil \cite{W} and De Donder \cite{DD} on the calculus
of variations in the 1930's.  In section \ref{fieldtheory}, I briefly
illustrate the main lines of this formulation using the example of a
scalar field, and I discuss its relation with the relativistic notions
of state and observable considered in \cite{c1}.  I then apply these
ideas to general relativity (GR) in Section \ref{GR}.

Unraveling the hamiltonian structure of GR has taken decades.  The
initial intricacies faced by Dirac \cite{dirac2} and Bergmann
\cite{bergmann} were slowly reduced in various steps, from the work of
Arnowit, Deser and Misner\cite{ADM} all the way to the Ashtekar
formulation \cite{ashtekar} and its variants.  Here, I discuss a far
simpler hamiltonian formulation of GR, constructed over a finite
dimensional configuration space and manifestly 4d generally covariant. 
The formulation is largely derived from the work of Esposito, Gionti
and Stornaiolo in \cite{gr4}.  (On covariant hamiltonian formulations
of GR, see also \cite{gr1,gr2,gr3,gr5,gr6,GRP}.)  I discuss two
interpretations of this formalism.  The first uses the coordinates,
while the second makes no direct reference to spacetime.  The four
spacetime coordinates drop out from the formalism (as the time
coordinate drops out from the ADM formalism) in an appropriate sense. 
I find this feature particularly attractive: the general relativistic
spacetime coordinates have no relation with observability and a
formulation of the theory in which they do not appear was long due.

I expect this formulation of GR to generalize to any matter coupling
and to any diffeomorphism invariant theory.  I think that it sheds
some light on the coordinate-independent physical interpretation of
the theory and helps clarifying issues that appear confused in the
hamiltonian formulations which are not manifestly covariant.  In
particular: what are ``states" and ``observables" in a theory without
background spacetime, without external time and without an asymptotic
region?  I close discussing the relevance of this analysis for the
problem of the interpretation of the formalism of quantum gravity. 
The formulation presented can be interpreted in terms of Toller's
reference system transformations \cite{Toller}, and provides a
physical interpretation for the spinnetwork to spinnetwork transition
amplitudes which can be computed in principle in loop quantum gravity
\cite{loop} and in the spinfoam models \cite{spinfoam}.

\section{Relativistic \\ hamiltonian  field theory}\label{fieldtheory}

There are several ways in which a field theory can be cast in
hamiltonian form.  One possibility is to take the space of the fields
at fixed time as the (nonrelativistic) configuration space $Q$.  This
strategy badly breaks special and, in a general covariant theory,
general relativistic invariance.  Lorentz covariance is broken by the
fact that one has to choose a Lorentz frame for the $t$ variable.  I
find far more disturbing the conflict with general covariance.  The
very foundation of general covariant physics is the idea that the
notion of a simultaneity surface all over the universe is devoid of
physical meaning.  Seems to me that it is better not to found
hamiltonian mechanics on a notion devoid of physical significance.

A second alternative is to formulate mechanics on the space of the
solutions of the equations of motion.  The idea goes back to Lagrange. 
In the generally covariant context, a symplectic structure can be
defined over this space using a spacelike surface, but one can show
that the definition is surface independent and therefore it is well
defined.  This strategy as been explored by Witten, Ashtekar and
several others \cite{GRP}.  The structure is viable in principle, but
very difficult to work with in practice.  The reason is that we do not
know the space of the solutions of an interacting theory.  Therefore
we must either work over a space that we can't even coordinatize, or
coordinatize the space with the initial data on some instantaneity
surface, and therefore, effectively, go back to the conventional fixed
time formulation.  Thus, this formulation has the merit of telling us
that the hamiltonian formalism is actually intrinsically covariant, but
it does not really indicate how to effectively deal with it in a
covariant manner.

The third possibility, which I consider here, is to use a covariant
finite dimensional space for formulating hamiltonian mechanics.  I
noticed in the companion paper \cite{c1} that in the relativistic
context the double role of the phase space, as the arena of mechanics
and the space of the states, is lost.  The space of the states, namely
the phase space $\Gamma$ is infinite dimensional in field theory,
virtually by definition of field theory.  But this does not imply that
the arena of hamiltonian mechanics has to be infinite
dimensional as well.  In particular, a main result of \cite{c1} is
that the natural arena for relativistic mechanics is the
extended configuration space $\cal C$ of the partial
observables.  Is the space of the partial observables of a field
theory finite or infinite dimensional?

Consider a field theory for a field $\phi(x)$ with $k$ components,
defined over spacetime $M$ with coordinates $x$, and taking values in
a $k$ dimensional target space $T$
\begin{eqnarray}
\phi: \ M &\longrightarrow & \ T \nonumber \\ 
x\ &\longmapsto& \phi(x). 
\label{function}
\end{eqnarray}
For instance, this could be Maxwell theory for the electric and
magnetic fields $\phi=(\vec E, \vec B)$, where $k=6$.  In order to
make physical measurements on the field described by this theory we
need $k$ measuring devices to measure the components of the field
$\phi$, and 4 devices (say one clock and three devices giving us the
distance from three reference objects) to determine the spacetime
position $x$.  Field values $\phi$ and positions $x$ are therefore the
partial observables of a field theory.  Therefore the operationally
motivated extended configuration space for a field theory is the
finite dimensional 4+$k$ dimensional space
\begin{equation}
  {\cal C}= M \times T.
\end{equation}
A correlation is  a point $(x,\phi)$ in $\cal C$. It represents a certain
value ($\phi$) of the fields at a certain spacetime point ($x$).  This
is the obvious generalization of the $(t,\alpha)$ correlations of the
pendulum of the example in \cite{c1}.

A motion is a physically realizable ensemble of correlations.  A
motion is determined by a solution $\phi(x)$ of the field equations. 
Such a solution determines a 4-dimensional surface in the (4+$k$
dimensional) space ${\cal C}$: the surface is the graph of the
function (\ref{function}).  Namely the ensemble of the points
$(x,\phi(x))$.  The space of the solutions of the field equations,
namely the phase space $\Gamma$, is therefore an (infinite
dimensional) space of 4d surfaces in the (4+$k$)-d configuration space
${\cal C}$.  Each state in $\Gamma$ determines a surface in $\cal C$.

Hamiltonian formulations of field theory defined directly on ${\cal
C}= M \times T$ are possible and have been studied.  The main reason
is that in a local field theory the equations of motion are local, and
therefore what happens at a point depends only on the neighborhood of
that point.  There is no need, therefore, to consider full spacetime
to find the hamiltonian structure of the field equations.  I refer the
reader to\cite{JK,HK} the ample references therein, and especially the
beautiful and detailed \cite{FTHM}.  What comes below is a very simple
and self-contained illustration of the formalism.

The difference with the finite dimensional case is that curves in
${\cal C}$ are replaced by 4d surfaces.  Thus, we need a hamiltonian
formalism determining these four dimensional surfaces in $\cal C$.  At
a point $p$ of ${\cal C}$, a curve has a tangent, leaving in
$T_{p}{\cal C}$, the tangent space of ${\cal C}$ at $p$.  A 4d surface
has four independent tangents $X_\mu$, or a ``quadritangent" $ X =
\epsilon^{\mu\nu\rho\sigma} X_\mu\otimes X_\nu
\otimes X_\rho \otimes X_\sigma $.

Consider a self interacting scalar field $\phi(x)$ defined on
Minkowski space $\cal M$, with interaction potential $V(\phi)$.  Its
field equation
\begin{equation}
\partial_{\mu}\partial^\mu\phi+ m^2\phi+V'(\phi) = 0 
\label{eq:fe}
\end{equation} 
can be derived from a hamiltonian formalism as follows.  To test the
theory (\ref{eq:fe}) we need {\em five\/} measuring devices: a clock
reading $x^0$, three devices that give us the spatial position $\vec
x$, and a device measuring the value of the field $\phi$.  Therefore,
the space $\cal C$ of the partial observables is the the five
dimensional space ${\cal C}={\cal M}\times T$, with coordinates
$(x,\phi)$.  Here $T=I\!\!  R$ is the target space of the field:
$\phi\in T$.  Let $\Omega$ be a 10d space with coordinates
$(x,\phi,p,p^\mu)$ carrying the the Poincar\'e-Cartan four-form
\begin{equation} 
    \theta = p\, d^4x + p^\mu\, d\phi\wedge d^3x_{\mu}. 
\label{eq:thetascalar} 
\end{equation} 
Here $d^4x\equiv dx^1\!\wedge\!dx^2\!\wedge\!dx^3\!\wedge\!  dx^4$ and
$d^3x_{\mu}\equiv d^4x(\partial_{\mu})$.  (Geometrically, $\Omega$ is
not the cotangent space of $\cal C$; it is a subspace of the bundle of
its four-forms, or a dual first jet bundle \cite{FTHM}.)  Consider
the constraint 
\begin{equation} H = p + ({\scriptstyle{1\over 2}}
p^\mu p_{\mu} + {\scriptstyle{1\over 2}}m^2\phi^2 + V(\phi)) = 0
\label{eq:Hscalar} 
\end{equation} 
on $\Omega$.  ($H$--$p$ is the DeDonder-Weyl hamiltonian function.) 
Let $\omega$ be the restriction of the five form $d\theta$ to the
surface $\Sigma$ defined by $H=0$.  Then the solutions of
(\ref{eq:fe}) are the orbits of $\omega$.  An orbit of $\omega$ is an
integral surface of its null quadrivectors.  That is, it is a
4d surface immersed in $\Sigma$ whose quadri-tangent $X$
satisfies 
\begin{equation} 
    \omega(X)=0.  
    \label{eq:omegaxfield} 
\end{equation}
For simplicity, let's say that this surface can be coordinatized by
$x^\mu$, that is, it is given by $(x, \phi(x),p^\nu(x))$. 
Then $\phi(x)$ is a solution of the field equations (\ref{eq:fe}). 
Thus, (\ref{eq:omegaxfield}) is equivalent to the field equations
(\ref{eq:fe}).  

A state determines a 4d surface $(x, \phi(x))$ in the extended
configurations space $\cal C$.  It represents a set of combinations of
measurements of partial observables that can be realized in nature. 
The phase space $\Gamma$ is the space of these states, and is infinite
dimensional.

In the classical theory, a state determines whether or not a certain
correlation $(x,\phi)$, or a certain set of correlations
$(x_{1},\phi_{1})\ldots(x_{n},\phi_{n})$, can be observed.  They can
be observed iff the points $(x_{i},\phi_{i})$ lie on the 4d
surface that represents the state.  Viceversa, the observation of a
certain set of correlations gives us information on the state: the
surface has to pass by the observed points.  In the quantum theory, a
state determines the probability amplitude of observing a correlation,
or a set of correlations.

Notice that there is an important difference between a finite
dimensional system and a field theory.  For the first, the measurement
of a finite number of correlations can determine the state.  In the
quantum theory, a single correlation may determine the state.  For
instance, if we have seen the pendulum in the position $\alpha$ at
time $t$, we then know the quantum state uniquely.  It follows that
quantum mechanics determines uniquely the probability amplitude
$W(\alpha',t';\alpha,t)$ for observing a correlation $(\alpha',t')$
after having observed a correlation $(\alpha,t)$.  Clearly
\begin{equation}
W(\alpha',t';\alpha,t)= \langle \alpha',t'| \alpha,t \rangle
    \label{Wp}
\end{equation}
where $|\alpha,t \rangle$ is the eigenstate of the Heisenberg operator
$\hat\alpha(t)$ with eigenvalue $\alpha$.  In field theory, on the
other hand, an infinite number of measurement is required in principle
to uniquely determine the state.  We can measure any finite number of
correlations, and still do not know the state.  Predictions in field
theory are therefore always given on the basis of some a priori
assumption on the state.  In quantum theory, this additional
assumption, typically, is that the field is in a special state such as
the vacuum.  Thus, a prediction of the quantum theory takes the
following form: if the system is in the vacuum state and we observe a
certain set of correlations 
$(x_{1},\phi_{1})\ldots (x_{n},\phi_{n})$, what is the probability
amplitude
\begin{equation}
W(x'_{1},\phi'_{1}\ldots x'_{n'},\phi'_{n'}; x_{1},\phi_{1}\ldots
x_{n},\phi_{n})
\label{W}
\end{equation}
of observing a certain other set $(x'_{1},\phi'_{1})\ldots
(x'_{n'},\phi'_{n'})$ of correlations ?  The quantities (\ref{W}) are
directly related to the usual $n$ point distributions of field theory. 
The relation is the same as the one that transforms the position basis
to the energy basis for an harmonic oscillator, that is, for instance
\begin{equation}
W(x',\phi';x,\phi)=\sum_{n,m} \overline{\psi_{n}(\phi')}\psi_{m}(\phi)
\ \langle n,x'| m,x \rangle
\label{W2}
\end{equation}
where $| n,x \rangle \sim (a^\dagger(x))^n | 0 \rangle $ is the state 
with $n$ particles in $x$. The distributional character of these
quantities will be studied elsewhere. 

For later comparison with GR, notice that the spacetime component
${\cal M}$ of the relativistic configuration space ${\cal C}={\cal
M}\times T$ is essential in the description, since the predictions of
the theory regard precisely the dependence of the partial observable
$\phi$ on the partial observables $x^\mu$.

The simplicity, covariance and elegance of this hamiltonian
formulation is quite remarkable.  I find it particularly attractive
from the conceptual point of view, because the notions of observable
and state on which it is based are operationally well founded,
relativistic and covariant.  I now apply these ideas to GR.

\section{Covariant hamiltonian GR}\label{GR}

GR can be formulated in tetrad-Palatini variables as follows.  I
indicate the coordinates of the spacetime manifold $M$ as $x^\mu$,
where $\mu=0,1,2,3$.  The fields are a tetrad field $e_{\mu}^I$ and a
Lorentz connection field $A_{\mu}^{IJ}$ (antisymmetric in $IJ$) where
$I=0,1,2,3$ is a 4d Lorentz index, raised and lowered with the
Minkowski metric.  The action is
\begin{equation} 
    S=\int 
e_{\mu}^{I}e_{\nu}^{J} F_{\rho\sigma}^{KL}\  
\epsilon^{\mu\nu\rho\sigma}\epsilon_{IJKL}\ 
d^4x \label{eq:S}
\end{equation} 
where $F_{\mu\nu}^{IJ}$ is the curvature of $A_{\mu}^{IJ}$.  The field
equations turn out to be 
\begin{equation}
D_{\nu}(e_{\rho}^{J}e_\sigma^{J})\epsilon^{\mu\nu\rho\sigma}= 0, \ \ \
e_\mu^{I}F_{\nu\rho}^{JK}\epsilon^{\mu\nu\rho\sigma}
\epsilon_{IJKL}=0.  
\label{eq:EE} 
\end{equation} where $D_{\mu}$ is the covariant derivative of the
Lorentz connection.  The first equation implies that $A_{\mu}^{IJ}$ is
the spin connection determined by the tetrad field.  Using this, the
second is equivalent to the Einstein equations.  That is, if
$(e_{\mu}^I(x) , A_{\mu}^{IJ}(x))$ satisfy (\ref{eq:EE}), then the
metric tensor $g_{\mu\nu}(x)=e_{\mu}^{I}(x)e_{\nu I}(x)$ satisfies the
Einstein equations.  I shall thus refer at (\ref{eq:EE}) as the
Einstein equations.  Below, these equations are derived from a
hamiltonian formalism, derived from \cite{gr4}. 

\subsection{First
version}\label{first}

Consider the (4+16+24) dimensional space $\tilde{\cal C}$ with
coordinates $(x^\mu, e_{\mu}^{I}, A_{\mu}^{IJ})$.  We have
${\tilde{\cal C}}=M\times T$, where $T$ is the target
space on which the tetrad-Palatini fields of GR take
value and $M$ is the spacetime manifold on which they are defined. 
\begin{equation} e^I=e^I_{\mu}\,
dx^\mu, \ \ \ \ A^{IJ}=A^{IJ}_{\mu}\, dx^\mu 
\label{eq:eA}
\end{equation} 
are one-forms on ${\tilde{\cal C}}$.  For any function or form on
${\tilde{\cal C}}$ with Lorentz indices, the Lorentz covariant
differential is defined by 
\begin{equation} 
    Dv^I = dv^I + A^I_J v^J.
\label{eq:D} 
\end{equation} 
I now introduce the main objects of the formalism: the
form on ${\tilde{\cal C}}$ 
\begin{equation}
\theta = {1\over 2}\ \epsilon_{IJKL}\ e^I \wedge e^J \wedge DA^{KL}
\label{eq:theta}
\end{equation} 
and the presymplectic form $\omega = d\theta$.  (Notice that $dA^{KL}
= d(A^{IJ}_{\mu}dx^\mu) = dA^{IJ}_{\mu}\wedge dx^\mu$, because
$A^{IJ}_{\mu}$ and $x^\mu$ are independent coordinates on
${\tilde{\cal C}}$.)  Now, the remarkable fact is that the
presymplectic form $\omega$ defines GR completely.  In
fact, its orbits, defined by
\begin{equation} 
    \omega(X) = 0 
    \label{eq:omegaXGR} 
\end{equation} 
where $X$ is the quadritangent to the orbit, are the solutions of the
Einstein equations.  That is: assuming for simplicity that the $x^\mu$
coordinatize the orbit, namely that the orbit is given by $(x^\mu,
e_\mu^{I}(x), A_{\mu}^{IJ}(x))$, then it follows from
(\ref{eq:omegaXGR}) that $(e_\mu^{I}(x), A_{\mu}^{IJ}(x))$ solve the
Einstein equations.  Viceversa, if $(e_\mu^{I}(x) , A_{\mu}^{IJ}(x))$
solve the Einstein equations, then $(x^\mu, e_\mu^{I}(x) ,
A_{\mu}^{IJ}(x))$ is an orbit of $\omega$.  Equation (\ref{eq:omegaXGR})
is equivalent to the Einstein equations.  The demonstration is a
straightforward calculation given in the Appendix.

\subsection{Second version}\label{second}

The simplicity of the formulation of GR described above is quite
striking.  I find the following observations even more remarkable. 
The space $\tilde{\cal C}$ contains the field variables $A_{\mu}^{IJ}$
and $e_{\mu}^{I}$ as well as the spacetime coordinates $x^\mu$.  Since
the theory is coordinate invariant, we are in a situation analogous to
the finite dimensional cases of the relativistic particle, or the
cosmological model, described in \cite{c1}.  In those examples, the
unphysical lagrangian evolution parameter $t$ dropped out of the
formalism; not surprisingly, since it had nothing to do with
observability.  Here, similarly, we should expect the coordinates
$x^\mu$ to drop out of the formalism.  In fact, it is well known that
the gauge invariant quantities of GR are coordinate independent.  They
are obtained by solving away the coordinates from quantities
constructed out of the fields.  Therefore the theory should actually
live on the sole field space $T$ with coordinates $A_{\mu}^{IJ}$ and
$e_{\mu}^{I}$, without reference to the spacetime coordinates $x^\mu$. 
Is this possible ?

The remarkable aspect of the expression (\ref{eq:theta}) of the form
$\theta$ is that the differentials of the spacetime coordinates
$dx^\mu$ appear only within the one-forms $e^I=e_\mu^{I}dx^\mu$ and
$A^{IJ}=A_\mu^{IJ}dx^\mu$.  This fact indicates that the sole role of
the $x^\mu$ is to arbitrarily coordinatize the orbits in the 40d space
of the fields $(e_\mu^{I}, A_{\mu}^{IJ})$. We can therefore
reinterpret the formalism of the previous section dropping the
spacetime part of $\tilde{\cal C}$.

Let $V$ be a 4d {\em vector\/} space.  Clearly $V$ is not spacetime;
it can be interpreted as a ``space of directions".  Let $\cal C$ be
the 40d space of the one-forms $(e^{I}, A^{IJ})$ on $V$.  Notice that
$\cal C$ is the space ${\cal C}=V^*\otimes {\cal P}$ of the 4d
one-forms with value in the algebra $P$ of the Poincar\'e group. 
Choosing a basis $a_\mu$ in $V$, the coordinates on $\cal C$ are
$(e_\mu^{I}, A_{\mu}^{IJ})$ and $\cal C$ can be identified with the
target space $T$ of the tetrad-Palatini fields.  Consider a 4d surface
immersed in $\cal C$.  The tangent space $T_p$ to this surface at a
point $p=(e^{I}, A^{IJ})$ is a 4d vector space.  This space can be
identified with $V$.  In particular, given an arbitrary choice of
coordinates $x^\mu$ on the surface, we identify the basis
$\partial_\mu$ of $T_p$ with the basis $a_\mu$ of $V$.  Therefore we
have immediately the ten one-forms $(e^{I}, A^{IJ})$ on the tangent
space $T_{p}$.  That is, $e^{I}(\partial_\mu)=e_\mu^{I}$ and
$A^{IJ}(\partial_\mu)=A_\mu^{IJ}$.

Consider now a form $\alpha=\alpha_I de^I+\alpha_{IJ} dA^{IJ}$ on
$\cal C$.  $\alpha$ is a one-form on $T_p$ (valued in the
one-forms over $\cal C$), and we can write
$\alpha(\partial_\mu)=\alpha_I de_\mu^I+\alpha_{IJ} dA_\mu^{IJ}$. But 
notice that $\alpha$ determines also a two-form on $T_p$ by 
\begin{equation} 
\alpha(\partial_\mu{\scriptstyle\otimes}\partial_\nu)= \alpha_I\,
\partial_\mu e_\nu^{I}+ \alpha_{IJ} \, \partial_\mu A_\nu^{IJ}.
\label{eq:dEdA} 
\end{equation} 
It follows that $\omega=d\theta$, with $\theta$ given by
(\ref{eq:theta}), acts on multivectors in $T_p$.  In particular, it
acts on the quadritangent $X=\epsilon^{\mu\nu\rho\sigma} 
\partial_{\mu}{\scriptstyle\otimes}\partial_{\nu}{\scriptstyle\otimes}
\partial_{\rho}{\scriptstyle\otimes}
\partial_{\sigma}$ of $T_p$.  The vanishing (\ref{eq:omegaXGR}) of
$\omega(X)$ is equivalent to the Einstein equations (see Appendix).

Therefore the theory is entirely defined on the 40d space $\cal C$. 
The states are the 4d surfaces in $\cal C$, whose tangents are in the
kernel of $d\theta$.  This is all of GR. The spacetime part $M$ of
$\tilde{\cal C}\sim M\times {\cal C}$ is eliminated from the
formalism.  The only residual role of the $x^\mu$ is to arbitrarily
parametrize the states, precisely as for the unphysical lagrangian
parameter in the examples of \cite{c1}.  Below I study the physical
interpretation of $\cal C$.

\section{Physical interpretation of $\cal C$} \label{ph}

\subsection{Classical theory: reference system transformations}

As discussed in \cite{c1}, in general, coordinates of the extended
configuration space ${\cal C}$ are the partial observables of the
theory.  A point in ${\cal C}$ represents a correlation between these
observables, that is, a possible outcome of a simultaneous
measurements of the partial observables, which give information on the
state of the system, or that can be predicted by the theory.  What are
the partial observables and the correlations of GR captured in the
space ${\cal C}$?  Can we give the space of the Poincar\'e algebra
valued 4d one-forms ${\cal C}$ a direct physical interpretation?

Assume that the measuring apparatus includes a reference system formed
by physical objects defining three orthogonal axes and a clock. 
Following Toller \cite{Toller}, we take a transformation $\cal T$ of
the Poincar\'e group (in the reference system) as the basic operation
defining the theory.  That is, assume that the basic operation that we
can perform is to displace the reference system by a certain length in
a spacial direction, or wait a certain time, or rotate it by a certain
amount, or boost it at a certain velocity.  The operational content of
GR can be taken to be the manner in which the transformations $\cal T$
fit together \cite{Toller}.  That is, we assume that a number of these
transformations, ${\cal T}_1,...,{\cal T}_n$, can be performed and
that it is operationally meaningful to say that two transformations 
${\cal T}_i$ and ${\cal T}_j$ start at the same reference system, or
arrive at the same reference system, or one starts where the other
arrives.  We identify one transformation as the measurement of a
partial observable (the value of the partial observable is given by
the Poincar\'e group element giving the magnitude of the
transformation).

A set of such measurements is therefore an oriented graph $\gamma$
where each link $l$ carries an element $U_{l}$ of the Poincar\'e group. 
Arbitrarily coordinatize the nodes of the graph with coordinates $y$. 
In the classical theory, we assume that arbitrarily many and
arbitrarily fine transformations can be done, so that we can take the
elements of the Poincar\'e group as infinitesimal, namely in the
algebra, and the coordinates $y$ as smooth.  An infinitesimal
transformation can therefore be associated to an infinitesimal
coordinate change $dy$.  As there is a 10d algebra of available
transformations, there is a 10d space of infinitesimal transformations
and the coordinates $y$ are ten dimensional.  However, it is an
experimental fact --coded in the theory-- that rotations and boosts
close and realize the Lorentz group.  That is, the relations between
sets of physical rotations or boosts are entirely determined
kinematically by the Lorentz group.  The same is not true for
displacements.  Therefore the space of the $y$'s is naturally fibrated
by 6-d fibers isomorphic to the Lorentz group \cite{Toller}. 
Coordinatize the remaining 4d space (the space of the fibers) with 4d
coordinates $x^\mu$.  It follows that the non-trivial infinitesimal
transformations assign an element of the Poincar\'e algebra $\cal P$ to an
infinitesimal displacement $dx^\mu$.  A single correlation is then the
determination of a Poincar\'e algebra element for each $dx^\mu$.  The
space of the correlations is therefore the space of the Poincar\'e
algebra valued 4d one-forms, which is precisely $\cal C$.

We can restrict the partial observables to a smaller number.  First,
we are using a first order formalism with configuration variables as
well as momenta.  Either the $e^I$ or the $A^{IJ}$ alone suffice to
characterize a solution.  With the first choice, we can take physical
lengths and angles associated to the $dx^\mu$ displacements as partial
observables.  With the second, we can take the Lorentz rotation part
of the transformations $\cal T$.  This gives infinitesimal Lorentz
transformations $R^{IJ}=A^{IJ}_\mu dx^\mu$ as partial observables.  We
can also exploit the internal Lorentz gauge invariance of the theory
to partially gauge fix the Lorentz group.  As well known, for
instance, $A^{IJ}$ can be gauge fixed to an element of $so(3)$. 
Finally, one can further gauge fix by solving explicitly the
dependence of some partial observables as functions of others; see for
instance \cite{GPS} for a realistic way of doing this.  I shall not
deal with this possibility here. 

\subsection{Quantum theory: spin networks}

In the quantum theory, quantum discreteness does not allow us to go to
the continuous description.  A finite set of partial measurements must
therefore be represented by the graph $\gamma$ with elements of the
Poincar\'e group $U_{l}$ associated the links $l$.  If we restrict to
configuration observables we can take the $U_{l}$ to be in the Lorentz
group, or, in gauge fixed form, in the rotation group.  In general,
quantum theory gives the probability amplitude to observe a certain
ensemble of partial observables, given that a certain other ensemble
of observables has been observed.  See Section IV of \cite{c1} and
\cite{rr}.  Therefore, we should expect that the predictions of a
quantum theory of GR can be cast in the form of probability amplitudes
$W(\gamma',U'_{l'};\gamma, U_{l})$.

Now, the quantities $W(\gamma',U'_{l'};\gamma, U_{l})$ are precisely
of the form spinnetwork to spinnetwork transition amplitudes which
can be computed, in principle, in loop quantum gravity (see
\cite{loop}, and references therein) and in the spinfoam models (see
\cite{spinfoam} and references therein).  More precisely, we can
write, in analogy with (\ref{W2})
\begin{equation}
W(\gamma',U'_{l'};\gamma, U_{l})= 
\sum_{j',j} \overline{\psi_{j'}(U'_{l'})}\ \psi_{k}(U_{l})
\ \langle \gamma',j'| \gamma, j \rangle
\label{Ws}
\end{equation}
where $j'$ (respectively $j$) represents the possible labels of a
spinnetwork with graph $\gamma'$ (respectively $\gamma$),
$\psi_{j}(U_{l})$ is the spinnetwork function on the group, and $
\langle \gamma',j'| \gamma, j \rangle$ is the (physical) spinnetwork to
spinnetwork transition amplitude.  See \cite{loop2} for details. 
Therefore the hamiltonian structure illustrated here provides a
conceptual framework for the interpretation of these transition
amplitudes.  Notice that no trace of position or time remains in these
expressions.

\section{Conclusion}

The shift in perspective defended in the companion paper \cite{c1} is
partially motivated by special relativity, but it is really forced by
general relativity.  The notion of initial data spacelike surface
conflicts with diffeomorphism invariance.  A generally covariant
notion of instantaneous state, or evolution of states and observables
in time, make little physical sense.  In a general gravitational field
we cannot assume that there exists a suitable asymptotic region, and
therefore we do not have a notion of scattering amplitude and $S$
matrix.  In this context, it is not clear what we can take as states
and observables of the theory, and what is the meaning of dynamics. 
In the paper \cite{c1} and in this paper I have attempted a
relativistic foundation of mechanics, that could provide clean notions
of states and observables, making sense in an arbitrary general
relativistic situation, as well as in quantum theory.

I have argued that mechanics can be seen as the theory of the
evolution in time only in the nonrelativistic limit.  In general,
mechanics is a theory of relative evolution of partial observables
with respect to each other.  More precisely, it is a theory of
correlations between partial observables.  Given a state, classical
mechanics determines which correlations are observable and quantum
mechanics gives the probability amplitude (or probability density) for
each correlation.

In this paper I have applied the ideas of \cite{c1} to field theory. 
I have argued that the relativistic notions of state and observable 
lead naturally to the  formulation of field theory over a finite
dimensional space.  The application of this formulation to general
relativity leads to a remarkably simple hamiltonian formulation, in
which the physical irrelevance of the spacetime coordinates becomes
manifest. 

General relativity can be formulated simply as the pair $({\cal
C},d\theta)$.  $\cal C$ is the 40d space of the Poincar\'e valued 4d
one-forms, and $\theta$ is given by (\ref{eq:theta}).  The orbits of
$d\theta$ in $\cal C$, solutions of equation (\ref{eq:omegaXGR}), are
the solutions of the Einstein equations and form the elements of the
phase space $\Gamma$.  This is all of general relativity.

The compactness and simplicity of this hamiltonian formalism is quite
remarkable.  Notice for instance that the target space $T$, 
the extended configuration space $\cal C$,
the space $\Omega$ that carries the presymplectic form defining the
theory and the constraint surface $\Sigma$ are all identified.  The
form $\theta$ codes the dynamics as well as the symplectic structure
of the theory.  

The disappearance of the spacetime manifold $M$ and its coordinates
$x^\mu$ --which survive only as arbitrary parameters on the orbits--
generalizes the disappearance of the time coordinate in the ADM
formalism and is analogous to the disappearance of the lagrangian
evolution parameter in the hamiltonian theory of a free particle
\cite{c1}.  It simply means that the general relativistic spacetime
coordinates are not directly related to observations.  The theory does
not describe the dependence of the field components on $x^\mu$, but
only the relative dependence of the partial observables of $\cal C$ on
each other.

We can give $\cal C$ a direct physical interpretation in terms of
reference systems transformations.  This interpretation is illustrated
in Section \ref{ph}.  In the quantum domain, it leads directly to the
spin-network to spin-network amplitudes computed in loop quantum
gravity.

\section*{Appendix}

Here we prove the claim that equations (\ref{eq:omegaXGR}) is equivalent
to the Einstein equations (\ref{eq:EE}). 
Let us first write  $\omega$ explicitly.  We have from (\ref{eq:theta})
\begin{eqnarray} 
    \theta &=& {1\over 2}\
\epsilon_{IJKL}\ e_{\mu}^I dx^\mu \wedge e_{\nu}^J dx^\nu \wedge
\nonumber \\ && \ \ (d A_\sigma^{KL}\wedge dx^\sigma + A_\rho^{KM}
dx^\rho\wedge A_{\sigma M}{}^L dx^\sigma) \nonumber \\ &=& {\scriptstyle
1\over 2}\ \epsilon^{\mu\nu\rho\sigma} \epsilon_{IJKL} \ \ e_{\mu}^I
e_{\nu}^J d A_\rho^{KL}\wedge dx_\sigma \nonumber \\ && +
\epsilon^{\mu\nu\rho\sigma} \epsilon_{IJKL} \ \ e_{\mu}^I e_{\nu}^J
A_\rho^{KM} A_{\sigma M}{}^L\ d^4x. 
\label{eq:tildeomega}
\end{eqnarray} 
It follows 
\begin{eqnarray} 
    \omega &=& d\theta = \epsilon^{\mu\nu\rho\sigma}
\epsilon_{IJKL} \ e_{\mu}^I de_{\nu}^J \wedge dA_\rho^{KL}\wedge
dx_\sigma \nonumber \\ && + \epsilon^{\mu\nu\rho\sigma} \epsilon_{IJKL}
\ d(e_{\mu}^I e_{\nu}^J A_\rho^{KM} A_{\sigma M}{}^L)\wedge d^4x.
\label{omegatildetutta} 
\end{eqnarray} 
Coordinatize an orbit with the  $x^\mu$.  The tangents are then 
\begin{equation} 
X_{\mu}={\partial\over\partial x^\mu}+ \partial_\mu e_\nu^I\,
{\partial\over\partial e_\nu^I}+ \partial_\mu A_\nu^{IJ}\,
{\partial\over\partial A_\nu^{IJ}}.  
\end{equation} 
The components of $ X =
\epsilon^{\mu\nu\rho\sigma} X_{\mu}{\scriptstyle\otimes} 
X_{\nu}{\scriptstyle\otimes}
X_{\rho}{\scriptstyle\otimes} X_{\sigma}$ that give a nonvanishing
contribution when contracting with $\omega$ are the ones with at least
two $\partial_{\mu}={\partial/\partial x^\mu}$ components.  These 
are (I leave the ${\scriptstyle\otimes}$ understood in the notation)
\begin{eqnarray} 
     X_{\mu}&=&\epsilon^{\mu\nu\rho\sigma}\ [ \
\partial_{\mu}\partial_{\nu}\partial_{\rho}\partial_{\sigma} +
\nonumber \\ && +
\partial_{\mu}\partial_{\nu}\partial_{\rho}(\partial_{\sigma}
e_\tau^I){\partial\over\partial e_\tau^I}+
\partial_{\mu}\partial_{\nu}\partial_{\rho}
(\partial_{\sigma}A_\tau^{IJ}){\partial\over\partial A_\tau^{IJ}} +
\nonumber \\ && +
\partial_{\mu}\partial_{\nu}(\partial_{\rho}e_\tau^I)
{\partial\over\partial e_\tau^I}
(\partial_{\sigma}A_\epsilon^{IJ}){\partial\over\partial
A_\epsilon^{IJ}} + \ldots \ ] 
\label{quadritangent} 
\end{eqnarray}
From (\ref{omegatildetutta}) and (\ref{quadritangent}), we obtain
\begin{equation} 
    \omega(X)= K^\mu_{I} de_\mu^I+ K^\mu_{IJ}
dA_\mu^{IJ}+ K_\mu dx^ \mu, 
\end{equation} where 
\begin{eqnarray}
K^\mu_{I} &=&\epsilon_{IJKL} \epsilon^{\mu\nu\rho\sigma}\
F_{\nu\rho}^{JK}e_{\sigma}^L, \nonumber \\ K^\mu_{IJ} &=&\epsilon_{IJKL}
D_{\nu}e_\rho^{K}e_{\sigma}^L \epsilon^{\mu\nu\rho\sigma}.
\label{K}
\end{eqnarray} 
while $K_\mu$ vanishes if $K^\mu_{I} $ and $K^\mu_{IJ}$ do.  It
follows immediately that $\omega(X)= 0$ give the
Einstein equations (\ref{eq:EE}).

The Einstein equations are obtained even more directly in the second
version of the formalism. Of the four $\partial_{\mu}$, three contract 
the three $e^I$ and $A^{IJ}$ forms, giving their components, and one 
contracts either $de_{\mu}^I$ or $dA_{\nu}^{IJ}$, leaving simply 
\begin{equation} 
    \omega(X)= K^\mu_{I} de_\mu^I+ K^\mu_{IJ} dA_\mu^{IJ} 
\end{equation} 
where  $K^\mu_{I} $ and $K^\mu_{IJ}$ are again given by (\ref{K}).

\end{document}